\documentclass[12pt]{article}

\usepackage[inline,shortlabels]{enumitem}
\usepackage{float}
\usepackage{mathtools}
\usepackage{graphicx}
\usepackage[round]{natbib}
\usepackage[margin=1in]{geometry}
\usepackage{tikz}
\usepackage[english]{babel}
\usepackage{longtable}
\usepackage{color}
\usepackage{amssymb,amsmath,amsthm}
\usepackage{multirow}
\usepackage[titletoc,title]{appendix}
\usepackage{authblk}
\usepackage{setspace}
\usepackage{dsfont}
\usepackage[OT1]{fontenc}
\usepackage{refcount}
\graphicspath{ {./figs/} }

\usepackage{nameref,zref-xr} 
\zxrsetup{toltxlabel} 

\usepackage{subfig}
\usepackage[inline]{trackchanges}
\addeditor{Ivan}
\addeditor{Nima}
\addeditor{Kara}

\newtheorem{theorem}{Theorem}
\AtEndDocument{\refstepcounter{theorem}\label{finalthm}}
\AtEndDocument{\refstepcounter{equation}\label{finaleq}}

{
  \theoremstyle{definition}
  
}
{
  \theoremstyle{definition}
  \newtheorem{assumptioniden}{}
}
{
  \theoremstyle{definition}
  
}

\newtheorem{lemma}{Lemma}
\AtEndDocument{\refstepcounter{lemma}\label{finallemma}}

\DeclareMathOperator{\expit}{expit}

 \DeclareMathOperator{\var}{\mathsf{Var}}

\renewcommand{\P}{\mathsf{P}}

\newcommand{\p}{\mathsf{p}}
\newcommand{\q}{\mathsf{q}}
\newcommand{\g}{\mathsf{g}}
\newcommand{\e}{\mathsf{e}}

\newcommand{\rr}{\mathsf{r}}

\newcommand{\indep}{\mbox{$\perp\!\!\!\perp$}} 
 \newcommand{\dd}{\mathrm{d}}

\newcommand{\one}{\mathds{1}}
 
\newcommand{\E}{\mathsf{E}}

\renewenvironment{proof}{{\it Proof }}{\qed \\}

\DeclarePairedDelimiterX{\norm}[1]{\lVert}{\rVert}{#1}

\pgfdeclarelayer{background}
\pgfsetlayers{background,main}
\usetikzlibrary{arrows,positioning}
\tikzset{
>=stealth',
punkt/.style={
rectangle,
rounded corners,
draw=black, very thick,
text width=6.5em,
minimum height=2em,
text centered},
pil/.style={
->,
thick,
shorten <=2pt,
shorten >=2pt,}
}
\newcommand{\Vertex}[2]
{\node[minimum width=0.6cm,inner sep=0.05cm] (#2) at (#1) {$\footnotesize#2$};
}
\newcommand{\Vertexr}[2]
{\node[rectangle, draw, minimum width=0.6cm,inner sep=0.05cm] (#2) at (#1) {$\footnotesize#2$};
}
\newcommand{\ArrowR}[3]%
{ \begin{pgfonlayer}{background}
\draw[->,#3] (#1) to[bend right=30] (#2);
\end{pgfonlayer}
}
\newcommand{\ArrowL}[3]%
{ \begin{pgfonlayer}{background}
\draw[->,#3] (#1) to[bend left=45] (#2);
\end{pgfonlayer}
}
\newcommand{\EdgeL}[3]%
{ \begin{pgfonlayer}{background}
\draw[dashed,#3] (#1) to[bend right=-45] (#2);
\end{pgfonlayer}
}

\newcommand{\Arrow}[3]%
{ \begin{pgfonlayer}{background}
\draw[->,#3] (#1) -- +(#2);
\end{pgfonlayer}
}
\newcommand{\titlepaper}{Efficient and flexible estimation of natural
  mediation effects under intermediate confounding and monotonicity
  constraints}

\date{\today}
\author[1]{Kara E. Rudolph\thanks{corresponding author:}}
\author[2]{Iv\'an D\'iaz}
\affil[1]{\small Department of
  Epidemiology, Mailman School of Public Health, Columbia University.}
\affil[2]{\small Division of Biostatistics, Weill Cornell Medicine.}

\usepackage[colorlinks,citecolor=blue,urlcolor=blue]{hyperref}
\title{\titlepaper}

\begin{document}
\maketitle

\begin{abstract}
Natural direct and indirect effects are mediational estimands that decompose the average treatment effect and describe how outcomes would be affected by contrasting levels of a treatment through changes induced in mediator values (in the case of the indirect effect) or not through induced changes in the mediator values (in the case of the direct effect). Natural direct and indirect effects are not generally point-identifiable in the presence of a treatment-induced confounder, however they may still be identified if one is willing to assume monotonicity between a treatment and the treatment-induced confounder. We argue that this assumption may be reasonable in the relatively common encouragement-design trial setting where intervention is randomized treatment assignment and the treatment-induced confounder is whether or not treatment was actually taken/adhered to. We develop efficiency theory for the natural direct and indirect effects under this monotonicity assumption, and use it to propose a nonparametric, multiply robust estimator.  We demonstrate the finite sample properties of this estimator using a simulation study, and apply it to data from the Moving to Opportunity Study to estimate the natural direct and indirect effects of being randomly assigned to receive a Section 8 housing voucher---the most common form of federal housing assistance---on risk developing any mood or externalizing disorder among adolescent boys—-possibly operating through various school and community characteristics.
\end{abstract}

\section{Introduction}
Researchers are frequently interested in summarizing mediational effects across individuals (or other units). For example, in trying to understand reasons underlying an unexpected treatment effect, it may be helpful to decompose the overall effect into the portion operating through some intermediate variables (i.e., mediators)---called the indirect effect---hypothesized to be responsible for the unexpected overall effect, and the portion not operating through those intermediate variables---called the direct effect. Natural direct and indirect effects (NDE, NIE, which we also refer to as natural (in)direct effects) are a type of mediation estimand that allows for such a decomposition. They are descriptive in nature \citep{pearl2013direct}; the NIE describes how individuals' outcomes would be affected by changes in their mediator values induced by contrasting levels of a treatment/exposure. (We formally define the NIE and NDE in terms of counterfactual notation in Section \ref{sec:not}.) 
  
Generally, the NDE/NIE are not point-identifiable in the presence of treatment-induced confounder of the mediator-outcome relationship \citep{avin2005identifiability}. To give intuition for this, we first define a counterfactual outcome as an outcome variable had treatment been set to some value $a$ and the mediator been set to some value, $m$, possibly contrary to fact, denoted $Y_{A=a, M=m}$ (we provide a more detailed description of notation in Section \ref{sec:not}). Similarly, we define a counterfactual mediator as a mediator variable had treatment been set to some value $a$, possibly contrary to fact, denoted $M_{A=a}$. NDE/NIE are then generally defined in terms of nested counterfactual outcomes, such as $Y_{A=a, M_{A=a'}}$ \citep{pearl2001direct}. Identification of the counterfactual outcome $Y_{A=a, M_{A=a'}}$ from observed data becomes a problem, because setting treatment to a certain value ($A=a$) induces a counterfactual treatment-induced confounder under the same treatment, denoted $Z_{A=a}$. Setting treatment to the other value in the counterfactual mediator ($A=a'$) induces a counterfactual treatment-induced confounder under the other value of treatment, denoted $Z_{A=a'}$. The two treatment-induced counterfactual confounders, $Z_{A=a}$ and $Z_{A=a'}$ are correlated with each other, because they share unmeasured
  common causes, $U_Z$.  $Z_{A=a}$ is also correlated with $Y_{A=a, M=m}$, and $Z_{A=a'}$ is also correlated with $M_{A=a'}$, which precludes identification.  

This general lack of identification in the presence of a treatment-induced confounder presents a problem for applied mediational research, because such variables are near-ubiquitous. For example, in trials where an individual or community is randomized to a particular treatment but cannot be forced to take the treatment, treatment take-up/ adherence represents a treatment-induced confounder. In observational data, there may be multiple intermediate variables linking treatment/exposure to the outcome, but only a subset are of interest to examine as mediators (for example, in the case discussed above where one is interested in examining which mediators contribute to the unexpected overall effect). The remaining intermediate variables would be considered as treatment-induced confouders. 

However, particular cases exist when the NDE and NIE are identifiable even in the presence of one or more treatment-induced confounders \citep{miles2022causal,tchetgen2014identification}. One of these cases is when one can assume monotonicity between a treatment and one or more binary treatment-induced confounders \citep{tchetgen2014identification}. Assuming monotonicity is also sometimes referred to as assuming ``no defiers'' \citep{Angristetal&Imbens&Rubin96}. In the context of an encouragement design intervention, this means that: if treatment assignment acts to increase the likelihoood treatment take-up, that there is no one for whom receiving treatment would result in them not taking the treatment, but not receiving the treatment would result in them taking the treatment. 

Monotonicity may seem like a restrictive assumption, 
but, in fact, it may be reasonable in some common scenarios. For example, consider the relatively common trial setting where the intervention is randomized treatment assignment and the treatment-induced confounder is whether or not treatment was actually taken. In this setting, we may feel comfortable assuming that treatment assignment would not make anyone more likely \textbf{not} to take the treatment. 
Indeed, monotonicity in this context is frequently assumed when using instrumental variables (IV) to identify causal effects \citep{Angristetal&Imbens&Rubin96}. Using IV as an identification strategy is common in the context of randomized trials where randomized assignment is the IV and treatment uptake or adherence is the exposure of interest. We consider this same context as our motivating example, but consider so-called intent-to-treat effects, where randomized assignment is our treatment of interest (not an IV), and treatment uptake/adherence is the treatment-induced confounder. 

Although monotonicity may be a reasonable assumption in 
trial settings, we know of no nonparametric estimators of natural (in)direct effects in this context (i.e., in the presence of treatment-induced confounders, assuming monotonicity). 
Consequently, we develop efficiency theory for the NDE/NIE under monotonicity, and use it to propose a nonparametric, multiply robust estimator of NDE/NIE based on solving the efficient influence function (EIF) estimating equation. Of relevance for real-world analyses, our estimator allows for continuous and/or multiple mediators and is cross-fitted. This estimator should be applicable whenever natural direct or indirect effects of treatment assignment are of interest in the presence potentially imperfect compliance acting as a single treatment-induced confounder. 

This paper is organized as follows. We give notation, the structural causal model, and review the definition and identification of NDE/NIE in Section 2.
We propose a nonparametric, robust estimator using 
flexible, data-adaptive regression methods in Section 3. Section 4 details a limited simulation study evaluating
the finite sample performance of the proposed estimator. In Section 5, we apply
the estimator to our motivating example estimating the NDE and NIE of being randomly assigned to receive a Section 8 housing voucher---the most common form of federal housing assistance---on risk developing any mood or externalizing disorder among adolescent boys----possibly operating through various school and community characteristics, using longitudinal data from the Moving to Opportunity Study (MTO) \citep{sanbonmatsu2011moving}. Section 6 concludes.

\section{Notation, structural causal model, estimand definition}
\label{sec:not}
Let $O=(W, A, Z, M, Y)$ denote the observed data, and let $O_1, \ldots, O_n$ denote a sample of $n$
i.i.d.~observations of $O$. 
Let $W$ denote a vector of observed baseline covariates, $W=f(U_W)$, where $U_W$ is unobserved exogenous error on $W$ and where function $f$ is assumed deterministic but unknown \citep{Pearl2009}. Let $A$ denote a binary treatment/exposure variable, $A=f(W,U_A)$. Let $Z$ denote single, binary treatment-induced confounder of the mediator-outcome relationship, e.g., initiation of treatment/ adherence to treatment assignment, $Z=f(W, A, U_Z)$, though we note that our results could be extended to accomodate multiple, binary $Z$. Let $M$ denote a set of mediating variables, $M=f(W, A, Z, U_M)$, that may be multiple, multi-valued, and/or continuous. Finally, let $Y$ denote a continuous or binary outcome, $Y=f(W,A,Z,M,U_Y).$ 

We use $\P$ to
denote the distribution of $O$, and $\P^c$ to denote the distribution
of $(O,U)$. We let $\P$ be an element of the nonparametric statistical
model defined as all continuous densities on $O$ with respect to some
dominating measure $\nu$. Let $\p$ denote the corresponding
probability density function, and $\mathcal{W, A, Z, M}$ denote the range
of the respective random variables. We let $\E$ and $\E^c$ denote
corresponding expectation operators, and define
$\P f = \int f(o)\dd \P(o)$ for a given function $f(o)$. We use
$\g(a \mid w)$ to denote the probability mass function of $A=a$
conditional on $W = w$, and $\e(a \mid m, z, w)$ to denote the
probability mass function of $A=a$ conditional on $(M, Z, W)=(m,z,w)$. We
use $\mu(m,z,a,w)$ to denote the outcome regression function
$\E(Y \mid M = m, Z = z, A=a, W = w)$.  
We use $\q(z \mid a,w)$ and $\rr(z \mid a,m,w)$ to denote the corresponding conditional densities
of $Z$.  

We define counterfactual variables in terms of interventions on the
nonparametric structural equation model (NPSEM). For simplicity of notation, we will use the random variable
with its corresponding intervention in the index to denote the
counterfactual. For example, $Y_{A=1}$ denotes the random variable
$f_Y(W, 1, Z, M, U_Y)$. In what follows, we drop the random variable from the index
of the counterfactual whenever the variable that is being intervened on is clear from context. For example, we simply use $Y_{a,m}$ to denote
$Y_{A=a,M=m}$, and $M_a$ to denote $M_{A=a}$. 

We are interested in the NDE and NIE, which decompose the overall average treatment effect (ATE) of the binary treatment/exposure as follows: 
\begin{equation*}
\E^c(Y_{1, M_1} - Y_{0, M_0}) =
      \underbrace{\E^c(Y_{1, M_1} - Y_{1, M_0})}_{\text{natural indirect
          effect (through $M$)}} +
  \underbrace{\E^c(Y_{1, M_0} - Y_{0,M_0})}_{\text{natural direct effect (not through $M$)}}
\label{eq:decomp}.
\end{equation*}

\citet{tchetgen2014identification} identified the NDE and NIE from observed data $O$ in the presence of treatment-induced confounders, $Z$. For a single, binary $Z$, the parameter $\E^c[Y_{a, M_{a'}}]$,
is identified as
\[\theta(a,a')=\theta_{1,1}(a,a')+\theta_{1,0}(a,a')+\theta_{0,0}(a,a'),\]
where
\begin{align*}
  \theta_{1,1}(a,a')&=\int \E(Y\mid a, m, 1, w)\dd\P(m\mid a', 1, w)\P(Z=1\mid
                a',w)\dd\P(w),\\
  \theta_{1,0}(a,a')&=\int \E(Y\mid a, m, 1, w)\dd\P(m\mid a', 0, w)[\P(Z=1\mid
                a,w)-\P(Z=1\mid a',w)]\dd\P(w),\\
  \theta_{0,0}(a,a')&=\int \E(Y\mid a, m, 0, w)\dd\P(m\mid a', 0, w)\P(Z=0\mid a,w)\dd\P(w),
\end{align*} under the following assumptions:

\begin{assumptioniden}[Monotonicity] If $a' < a,$ then $Z_{a'} \le Z_a$ for all $i$;\label{ass:mono}
\end{assumptioniden}
\begin{assumptioniden}[Sequential Exchangeability] $Y_{A=a,M=m}\indep A\mid W$,  $Y_{A=a,M=m}\indep M\mid W,A,Z$, and $M_{A=a}\indep A\mid W$;
\label{ass:exch}
\end{assumptioniden}
\begin{assumptioniden}[Positivity of treatment/exposure, treatment-induced confounder, and mediator mechanisms] \label{ass:pos} Assume:
  \begin{itemize}
      \item $\p(w)>0$ implies $\p(a\mid w)>0$ for $a\in\{0,1\}$;
      \item $[\P(Z=1 \mid a',w)\times I(z=z'=1) + \{\P(Z=1 \mid a, w) - \P(Z=0 \mid a',w)\}\times I(z=1, z'=0) + \P(Z=0 \mid a,w)\times I(z=z'=0)] >0 $ implies $\p(m \mid z', a', w)>0 $ and $\p(z \mid a, w)>0$ for $(z, z') \in \{(0,0), (1,0), (1,1)\}$; and
      \item $\p(m \mid z', a', w)>0$ implies $\p(m \mid z, a, w)>0$ for all observed $m, z, z'$ and for $(a, a') \in \{(0,0), (1,0), (1,1)\}$.
  \end{itemize}
\end{assumptioniden}

\section{Estimation}
We propose a one-step estimator for the statistical parameter $\theta(a,a')$ 
that is the sample average of the estimated uncentered efficient influence function (EIF). An R package to implement this estimator is included \url{link blinded for review}. The EIF for this parameter is given by the following expressions. Define
{\small\begin{align*}
  H_{Y,1,1} &= \frac{\one\{Z=1, A=a\}}{\P(a'\mid W)}\frac{\P(a'\mid
                M, 1, W)}{\P(a\mid M, 1, W)}\\
  H_{Y,1,0} &= \frac{\one\{Z=1, A=a\}}{\P(a'\mid W)\P(Z=0\mid a',W)}\frac{\P(a'\mid
                M, 1, W)}{\P(a\mid M, 1, W)}\frac{\P(Z=0\mid
                M,a',W)}{\P(Z=1\mid M,a',W)}
              \{\P(Z=1\mid
                a,w)-\P(Z=1\mid a',w)\}\\
  H_{Y,0,0} &= \frac{\one\{Z=0, A=a\}}{\P(a\mid W)}\frac{\P(a'\mid
                M, 0, W)}{\P(a\mid M, 0, W)}\frac{\P(Z=0\mid a,
                W)}{\P(Z=0\mid a',W)}\\
  H_{M,1,1}&=\frac{\one\{A=a', Z=1\}}{\P(a'\mid W)}\\
  H_{M,1,0}&=\frac{\one\{A=a',Z=0\}}{\P(Z=0\mid a',W)\P(a'\mid W)}\{\P(Z=1\mid
              a,w)-\P(Z=1\mid a',w)\}\\
  H_{M,0,0}&=\frac{\one\{A=a',Z=0\}}{\P(Z=0\mid a',W)\P(a'\mid W)}\P(Z=0\mid
              a,w)\\
  H_{Z,1,1}&=\frac{\one\{A=a'\}}{\P(a'\mid W)}\int\E(Y\mid a, m, 1,
              W)\dd\P(m\mid a',1,W)\\
  H_{Z,1,0}&=\left(\frac{\one\{A=a\}}{\P(a\mid W)}-\frac{\one\{A=a'\}}{\P(a'\mid W)}\right)\int\E(Y\mid a, m, 1,
              W)\dd\P(m\mid a',0,W)\\
  H_{Z,0,0}&=-\frac{\one\{A=a\}}{\P(a\mid W)}\int\E(Y\mid a, m, 0,
              W)\dd\P(m\mid a',0,W)\\
  H_{W,1,1}&=\int \E(Y\mid a, m, 1, w)\dd\P(m\mid a', 1, w)\P(Z=1\mid
              a',w)\\
  H_{W,1,0}&=\int \E(Y\mid a, m, 1, W)\dd\P(m\mid a', 0, W)[\P(Z=1\mid
              a,W)-\P(Z=1\mid a',W)]\\
  H_{W,0,0}&=\int \E(Y\mid a, m, 0, W)\dd\P(m\mid a', 0, W)\P(Z=0\mid
              a,W)
\end{align*}}

Then, for $(z,z')\in\{(1,1), (1,0), (0,0)\}$, the EIF of $\theta_{z,z'}(a,a')$
is equal to $\bar D_{z,z'}(O;\eta) = D_{z,z'}(O;\eta)-\theta_{z,z'}(a,a')$, where
\begin{align*}
  D_{z,z'}(O) &= H_{Y,z,z'}\left[Y - \E(Y\mid A,M,Z,W)\right]\\
             &+H_{Z,z,z'}[Z-\P(Z=1\mid A, W)]\\
             &+H_{M,z,z'}\left[\E(Y\mid a, M, z,
               W) - \int \E(Y,a,m,z,W)\dd\P(m\mid a',z',W)\right]\\
             &+H_{W,z,z'}.
\end{align*}
The EIF of $\theta(a,a')$ is then $\bar D(O) = D(O) - \theta(a,a')$, where $D(O) = D_{1,1}(O) + D_{1,0}(O) + D_{0,0}(O).$


Let $\rho$ denote $\int \E(Y\mid a,m,z,w)\dd \P(m\mid a',z,w)$. Let $\eta = (\g, \q, \e, \rr, \mu, \rho)$, where each is defined as in Section 2. Let $\hat\eta$ denote an estimator of $\eta$. Let $D(O,\P)=D(O,\eta),$ as $\eta$ contains all the relevant features of $\P$. Let 
$D(O,\hat{\eta})$ denote an estimator of $D(O,\eta)$. The one-step estimator we propose is the sample average of $D(O,\hat{\eta}).$
    
We use a cross-fitted version of this estimator. Cross-fitting is a data-splitting technique that weakens some of the technical assumptions (i.e., Donsker-type assumptions) required for asymptotic normality \citep{klaassen1987consistent,zheng2011cross, chernozhukov2016double}. We perform crossfitting for estimation of all the components of $\eta$ as follows. Let ${\cal V}_1, \ldots, {\cal V}_J$
denote a random partition of data with indices $i \in \{1, \ldots, n\}$ into $J$
prediction sets of approximately the same size such that 
$\bigcup_{j=1}^J {\cal V}_j = \{1, \ldots, n\}$. For each $j$,
the training sample is given by
${\cal T}_j = \{1, \ldots, n\} \setminus {\cal V}_j$. 
$\hat \eta_{j}$ denotes the estimator of $\eta$, obtained by training
the corresponding prediction algorithm using only data in the sample
${\cal T}_j$, and $j(i)$ denotes the index of the
validation set which contains observation $i$. We then use these fits, $\hat\eta_{j(i)}(O_i)$ in computing each efficient influence function, i.e., we compute $D(O_i, \hat\eta_{j(i)})$. 

Thus, our one-step estimator of $\theta(a,a')$ can be calculated in the following steps:
\begin{enumerate}
    \item  Let the components of $\eta$ be defined as above. With the exception of $\rho$, each can be estimated by cross-fitting a regression of the dependent variable on the independent variables and generating predicted probabilities (if the dependent variable is binary) or predicted values otherwise, setting the values of independent variables where indicated. For example, $\hat\g(a' \mid w)$ can be estimated by cross-fitting a logistic regression model of $A$ on $W$ and generating predicted probabilities that $A=a'$ for all observed $w$.  One could also use machine learning in model fitting, which is what we do in the simulations and data analysis.
    \item To estimate $\rho$,
    we use the predicted values from $\mu(a, m, z, W)$ as a pseudo-outcome in a new regression on variables $A, Z, W$, and then generate predicted values setting $A=a'$ and  $Z=z'.$ 
    \item The estimator of $\theta(a, a')$ is $\hat\theta(a,a')=\frac{1}{n}\sum_{i=1}^n D(O_i, \hat{\eta}_{j(i)}).$ 
    \item The variance can be estimated as the sample variance of $D(O, \hat{\eta}).$
\end{enumerate}

Let $\tilde \eta$ denote the probability limit of the estimator $\hat\eta$ in $L_2(\P)$ norm. The above estimator is multiply robust such that $\theta(a, a')$ is expected to be estimated consistently under one of the following: 
\begin{itemize}
    \item $(\tilde\mu, \tilde\rho, \tilde\q) = (\mu, \rho, \q) $ or
    \item $(\tilde\mu, \tilde\rho, \tilde\g) = (\mu, \rho, \g) $ or
    \item $(\tilde\mu, \tilde\g, \tilde\q) = (\mu, \g, \q) $ or
    \item $(\tilde\g, \tilde\e, \tilde\q, \tilde\rr) = (\g, \e, \q, \rr).$
\end{itemize}
The monotonicity assumption upon which this identification rests may be most likely tenable in encouragement-design randomized trials like the MTO study used in the illustrative application. In that case, estimation of $\g$ may not be necessary, so one would expect consistent estimation if either $(\mu, \q)$ or $(\mu, \rho)$ or $(\e, \q, \rr)$ were consistently estimated.
These robustness conditions are a consequence of the following Lemma, the proof of which is given in the appendix. 
\begin{lemma}
\label{lem:robust}
Let $D_{z,z'}(O;\tilde \eta)$ denote the EIF evaluated at a given value $\tilde\eta$ of the nuisance parameters. Then we have
\[\theta_{z,z'}(a,a')(\tilde\eta) - \theta_{z,z'}(a,a')(\eta) = -\E[\bar D_{z,z'}(O;\tilde \eta)] + R_{z,z'}(\eta,\tilde\eta),\]
where $R_{z,z'}(\eta,\tilde\eta)$ is a second order term given by:
\begin{align*}
R_{z,z'}(\eta,\tilde\eta) &=\E\big\{[\tilde H_{Y,z,z'} - H_{Y,z,z'}][\E(Y\mid A,M,Z,W)-\tilde \E(Y\mid A,M,Z,W)]\big\}\\ 
&+\E\left\{[\tilde H_{M,z,z'} - H_{M,z,z'}]\int \tilde \E(Y\mid a,M,z,W)[\dd \P(m\mid a', z', W) - \dd \tilde \P(m\mid a', z', W)]\right\}\\ 
&+\E\big\{[\tilde H_{Z,z,z'} - H_{Z,z,z'}][\P(Z=1\mid A,W)-\tilde \P(Z=1\mid A,W)]\big\}\\ 
&+ R^\star_{z,z'}(\eta,\tilde\eta),
\end{align*}
where, for each $(z,z')$, $R^\star_{z,z'}(\eta,\tilde\eta)$ is equal to:
{\small
\begin{align*}
    R^\star_{1,1}(\eta,\tilde\eta) 
    &= \int [\E(Y\mid a,m,1,w)-\tilde \E(Y\mid a,m,1,w)][\P(Z=1\mid a', w)-\tilde\P(Z=1\mid a', w)]\dd\tilde\P(m\mid a',1,w)\dd\P(w)\\
    &+\int \E(Y\mid a,m,1,w)[\P(Z=1\mid a', w)-\tilde\P(Z=1\mid a', w)][\dd\P(m\mid a',1,w)-\dd\tilde\P(m\mid a',1,w)]\dd\P(w)\\
    R^\star_{1,0}(\eta,\tilde\eta) 
    &= \int [\E(Y\mid a,m,1,w)-\tilde \E(Y\mid a,m,1,w)][\q_Z(w)-\tilde\q_Z(w)]\dd\tilde\P(m\mid a',0,w)\dd\P(w)\\
    &+\int \E(Y\mid a,m,1,w)[\q_Z(w)-\tilde\q_Z(w)][\dd\P(m\mid a',0,w)-\dd\tilde\P(m\mid a',0,w)]\dd\P(w)\\
    R^\star_{0,0}(\eta,\tilde\eta) 
    &= \int [\E(Y\mid a,m,0,w)-\tilde \E(Y\mid a,m,0,w)][\P(Z=1\mid a', w)-\tilde\P(Z=0\mid a', w)]\dd\tilde\P(m\mid a',0,w)\dd\P(w)\\
    &+\int \E(Y\mid a,m,0,w)[\P(Z=0\mid a', w)-\tilde\P(Z=0\mid a', w)][\dd\P(m\mid a',0,w)-\dd\tilde\P(m\mid a',0,w)]\dd\P(w),
\end{align*}}
for $\q_Z(w) = \P(Z=1\mid a,w) - \P(Z=1\mid a',w)$.
\end{lemma}

Inspection of the remainder terms reveals that the one-step estimator should be consistent in the configurations outlined above, as in such cases all the remainder terms become null. In addition, this lemma will allow us to prove an asymptotic normality result stating that the estimator converges in distribution to a normal random variable with variance equal to the non-parametric efficiency bound. Importantly, this result holds even when the nuisance parameters $\eta$ are estimated using using flexible regression techniques such as those available in the machine learning literature, as long as the regressions are cross-fitted as detailed above. The only requirement is that the second order terms of the previous lemma converge to zero in probability at rate $n^{-1/2}$. This can occur, for example, if all the nuisance regressions are estimated consistently at rate $n^{-1/4}$. This rate is much slower than the convergence rate of parametric models and is achievable under a flexible regression framework.

\begin{theorem}[Asymptotic normality of estimators]
Let $R = R_{1,1} + R_{1,0} + R_{0,0}$. Assume $R(\eta, \hat\eta)=o_P(n^{-1/2})$. Let $\hat H_{Y,z,z'}$, $\hat H_{M,z,z'}$, $\hat H_{Z,z,z'}$ denote the estimates of $H_{Y,z,z'}$, $H_{M,z,z'}$, $H_{Z,z,z'}$ constructed by plugging in estimates $\hat\eta$. Assume that there exists a constant $c$ such that $\P(\hat H_{Y,z,z'} < c) = \P(\hat H_{M,z,z'} < c) = \P(\hat H_{Z,z,z'} < c) = 1$. Then we have 
\[\hat\theta(a,a') -\theta(a,a') = \frac{1}{n}\sum_{i=1}^n \bar D(O_i,\eta) + o_P(n^{-1/2}).\]
\end{theorem}
The proof of this result for estimators that allow a second order expansion as in Lemma~\ref{lem:robust} follows standard empirical process theory and is given, for example, in \cite{diaz2021nonparametric}. 

This result together with the central limit theorem shows that $n^{1/2}(\hat\theta(a,a') -\theta(a,a'))$ converges to a normal distribution with variance $\var[\bar D(O,\eta)]$ equal to the non-parametric efficiency bound. Furthermore, this result is useful to compute standard errors and Wald-type confidence intervals for contrasts of the parameter $\theta(a,a')$ for varying values of $(a,a')$, by simply applying the Delta method to the corresponding contrast. 

\section{Simulation}
We conducted a limited simulation study to: 1) verify that our programmed estimator had the theoretical properties of: a) consistency under the robustness conditions established in Lemma \ref{lem:robust}, b) efficiency, and c) confidence interval coverage; and 2) illustrate the estimator's finite sample performance. We caution that these simulations should not be taken as representative of the estimator's general performance. 

We considered the following data-generating mechanism:
{\footnotesize
  \begin{align*}
    P(W_1 = 1) &= 0.6               \\
    P(W_2 = 1) &= 0.3               \\
    P(W_3 = 1 \mid W_1, W_2) &= \min(0.2 + 0.33(W_1 + W_2),1)\\
    P(A = 1) &= 0.5   \\
    P(Z = 1 \mid A, W) &= \expit(-\log(1.3) \times (W_1 + W_2 + W_3)/3 + 2A -1 ) \\
    P(M = 1 \mid Z, W) &= \expit(-\log(1.1)W_3 + 2Z - 0.9)                        \\
    P(Y = 1 \mid M, Z, W) &= \expit(-\log(1.3)\times (W_1 + W_2 + W_3)/3 + Z + M )        
  \end{align*}}

This DGM was constructed to align with the illustrative application in which: $A$ is randomly assigned and adheres to the exclusion restriction assumption by only having an effect on $M$ and $Y$ through $Z$ \citep{Angristetal&Imbens&Rubin96}, and $Z$ is monotonic with respect to $A$. 

We conducted 1,000 simulations for sample sizes $n \in \{1,000, 10,000\}$ and under correct specification of nuisance parameters in $\eta$ and under several misspecifications of parameters in $\eta$: 1) where $(\g, \e, \q, \rr)$ are correct but $(\mu, \rho)$ are misspecified, 2) where $(\mu, \rho, \g)$ are correct but $(\e, \q, \rr)$ are misspecified, and 3) where $(\mu, \rho, \q)$ are correct but $(\g, \e, \rr)$ are misspecified. We would still expect consistent estimation of $\theta(a,a')$ under each of these misspecifications given the robustness results that are a consequence of Lemma \ref{lem:robust}. Also, note that $A$ is exogenous in this simulation, so although we estimate $\g$ in this simulation, its estimation is not necessary.

We fit parameters with the highly adaptive lasso \citep{hejazi2020hal9001,benkeser2016highly,van2017generally} for the correctly specified scenarios, and fit parameters using an intercept-only model for the incorrectly specified scenarios.

We considered estimator performance in terms of absolute bias, absolute bias scaled by $\sqrt{n}$, influence curve-based standard error relative to the Monte Carlo-based standard error, standard deviation of the estimator relative to the efficiency bound scaled by $\sqrt{n}$, mean squared error relative to the efficiency bound scaled by $n$, and 95\% confidence interval (CI) coverage. 

Table \ref{tab:simdirect} shows simulation results for the natural direct effect, and Table \ref{tab:simindirect} shows simulation results for the natural indirect effect under correct specification of all nuisance parameters and various misspecifications, comparing sample sizes of 10,000 and 1,000. We expect consistent estimation in all scenarios. Focusing on the sample size of 10,000 and correct specification of all nuisance parameters, we see very little bias; relative standard error, relative standard deviation, and relative mean squared error close to 1; and 95\% confidence interval coverage close to 95\% for both the natural direct effect and natural indirect effect. For this same sample size of 10,000 but misspecification of some parameters, we maintain relatively consistent estimation, with bias increasing only slightly. However, the relative standard error, relative standard deviation, and relative mean squared error decrease markedly in both the $(\mu, \rho, g)$ and $(\mu, \rho, q)$ correct specifications, particularly in the case of the natural indirect effect. The 95\% confidence interval coverage also decreases under these misspecified scenarios, particularly again, for the natural indirect effect. In fact, the 95\% CI covers only 46\% of the time in the $(\mu, \rho, g)$ correct scenario for the natural indirect effect. The results under sample size of 1,000 are similar, albeit slightly worse on average, compared to the larger sample size. 

\begin{table}
\centering
\caption{Simulation results for the natural direct effect, $\theta(1,0) - \theta(0,0) = 0.1036$, efficiency bound= 1.7858. Abbreviations: relse = influence curve-based standard error relative to the Monte-Carlo-based standard error, relsd = standard deviation of the estimator relative to the efficiency bound scaled by $\sqrt{n}$, relrmse = mean squared error relative to the efficiency bound scaled by $n$, 95\%CI Cov = 95\% confidence interval coverage.}
\begin{tabular}{|p{4cm} |  p{1.2cm} p{1.4cm} p{1.2cm} p{1.2cm}   p{1.2cm} p{2.4cm}   | }
\hline \label{tab:simdirect}
Correctly specified nuisance parameters & $|\text{bias}|$& $\sqrt{n}|\text{bias}|$ & relse & relsd & relrmse &95\%CI Cov \\ 
 \hline
  \multicolumn{7}{|l|}{N=10,000}\\ \hline
All & 0.0004 & 0.0406 &  0.972 &  0.982 &  0.982  &  0.945\\
$\g, \e, \q, \rr$ & 0.0012  &0.124 & 1.01 & 0.964&   0.968  &  0.959\\
$\mu, \rho, \g$ &  0.0011 & 0.112 & 0.806& 0.857 &  0.861    & 0.885\\ 
$\mu, \rho, \q$ & 0.0010 & 0.100 & 0.966& 0.879 &  0.882  &  0.938\\
\hline \multicolumn{7}{|l|}{N=1,000}\\ \hline
All & 0.0019 & 0.0592 & 0.9302 & 0.9737 & 0.9743 & 0.9273 \\ 
$\g, \e, \q, \rr$  & 0.0047 & 0.1477 & 0.9941 & 0.9339 & 0.9400 & 0.9530 \\ 
$\mu, \rho, \g$  & 0.0018 & 0.0567 & 0.7962 & 0.8743 & 0.8749 & 0.8768 \\ 
$\mu, \rho, \q$ & 0.0070 & 0.2199 & 0.9449 & 0.8957 & 0.9103 & 0.9303 \\    \hline
\end{tabular}
\end{table}

\begin{table}
\centering
\caption{Simulation results for the natural indirect effect, $\theta(1,1) - \theta(1,0) = 0.0827$, efficiency bound= 0.9293. Abbreviations: relse = influence curve-based standard error relative to the Monte-Carlo-based standard error, relsd = standard deviation of the estimator relative to the efficiency bound scaled by $\sqrt{n}$, relrmse = mean squared error relative to the efficiency bound scaled by $n$, 95\%CI Cov = 95\% confidence interval coverage.}
\begin{tabular}{|p{4cm} |  p{1.2cm} p{1.4cm} p{1.2cm} p{1.2cm}   p{1.2cm} p{2.4cm}   | }
\hline \label{tab:simindirect}
Nuisance Parameters Correctly Specified & $|\text{bias}|$& $\sqrt{n}|\text{bias}|$ & relse & relsd & relrmse &95\%CI Cov \\ 
 \hline
  \multicolumn{7}{|l|}{N=10,000}\\ \hline
Correct   &    0.0004 & 0.0412 & 0.951 & 0.974  & 0.975   & 0.932\\
$\g, \e, \q, \rr$&  0.0012 & 0.116 & 0.988 & 0.916 &  0.924&    0.95\\
$\mu, \rho, \g$ & 0.0012 & 0.120 & 0.314 & 0.727 &  0.738  &  0.464\\
$\mu, \rho, \q$ & 0.0010 & 0.0993& 0.634 & 0.776 &  0.782 &   0.791 \\
\hline \multicolumn{7}{|l|}{N=1,000}\\ \hline
Correct & 0.0021 & 0.0672 & 0.8628 & 0.9743 & 0.9763 & 0.9010 \\ 
$\g, \e, \q, \rr$ &  0.0038 & 0.1200 & 0.9154 & 0.8786 & 0.8870 & 0.9270 \\ 
$\mu, \rho, \g$ & 0.0027 & 0.0867 & 0.3220 & 0.7412 & 0.7463 & 0.4859 \\ 
$\mu, \rho, \q$ &  0.0067 & 0.2131 & 0.6157 & 0.7922 & 0.8222 & 0.7444 \\    \hline
\end{tabular}
\end{table}

\section{Empirical Illustration}

Next, we apply our proposed estimators to estimate natural direct and indirect effects among adolescent boys whose families participated in the Moving to Opportunity Study (MTO) \citep{sanbonmatsu2011moving}. Briefly, MTO randomized families living in public housing in five cities in the United States (US), who volunteered to participate, to either receive a Section 8 housing voucher or not. (Families were actually randomized to one of three groups---two of which received a housing voucher, but we collapse these for simplicity, as others have done \citep{rudolph2021helped,rudolph2018mediation,rudolph2017composition,osypuk2012gender}.) Section 8 housing vouchers are the primary form of federal housing assistance, and they serve to subsidize rents on the private rental market for low-income households, thus making it more financially viable for a family to move out of public housing. In this example, we are interested in decomposing the average effect of being randomized to receive a Section 8 housing voucher ($A$) in early childhood on 10-15-year risk of developing any psychiatric disorder by adolescence ($Y$) (defined using the Diagnostic and Statistical Manual, Version IV (DSM-IV) criteria, as measured by the CIDI-SF \citep{kessler1998world,kessler2004world}) into the portion that operates through changes in the school environment and residential and school instability ($M$)---the natural indirect effect---and the portion that does not---the natural direct effect. 

MTO is an example of an encouragement intervention, because receipt of a Section 8 housing voucher (the randomized intervention) encourages families to move out of public housing and into a rental on the private market by subsidizing their rent. In the context of an encouragement intervention study design, the monotonicity assumption is reasonable. Randomized receipt of the encouragement---in this case, the housing voucher---should only serve to increase (as opposed to increasing for some and decreasing for others) the likelihood of complying with the intervention---in this case, moving within the first year to a lower poverty neighborhood ($Z$). 

In this example, we only consider boys whose families participated in MTO, as there have been marked differences in MTO's health effects between girls and boys in which being in the intervention group resulted in higher rates of post-traumatic stress disorder, mood disorder, any psychiatric disorder, smoking, and problematic drug use for boys, but not for girls \citep{clampet2011moving, sanbonmatsu2011moving,schmidt2017adolescence,kessler2014associations,rudolph2018mediation,kling2007experimental}. We also exclude the Baltimore site, because of evidence that the intervention was meaningfully different in this city versus the others \citep{rudolph2018mediation}, likely due to concurrent housing related-interventions in that city. These restrictions resulted in a rounded sample size of N=2,100. We adjust for numerous baseline covariates, $W$, which we detail in the appendix. We provide additional detail of mediator variables $M$ in the appendix as well. For simplicity, we used one imputed dataset, which was imputed using multiple imputation by chained equations \citep{buuren2010mice}. The mediator of school rank had the most missingness at 12\%, other mediators had 8-9\% missingness or no missingness. The outcome of any DSM-IV disorder had 8\% missingness. The randomized voucher assignment variable and moving variable had no missingness. Two baseline covariates had 2\% missing data (race/ethnicity and baseline neighborhood poverty), and the rest had no missing data.

We estimated the natural direct and indirect effects of being randomized
    to the Section 8 voucher group on risk of having a psychiatric disorder disorder in
    adolescence, 10-15 years later, mediated through features of the neighborhood and
    school environments (in the case of the indirect effect), and not (in the case of the direct effect), using the cross-fitted one-step estimator proposed here, with 2 folds. We used the SuperLearner ensemble method of combining machine learning algorithms in fitting the nuisance parameters, implemented with the \url{sl3} package \citep{coyle2021sl3-rpkg}; this approach weights the algorithms to minimize the 10-fold cross-validated prediction error \citep{van2007super}. We included the following algorithms: intercept-only regression, generalized linear regression, lasso \citep{tibshirani1996regression}, and gradient boosted machines \citep{chenXGBoost}. Columbia University determined this analysis of deidentified data to be non-human subjects research.
    
\begin{figure}
  \caption{Natural direct and indirect effects estimates of being randomized
    to the Section 8 voucher group on risk of  having a psychiatric disorder disorder in
    adolescence, 10-15 years later, mediated through features of the neighborhood and
    school environments, among boys. Estimates and 95\% CIs. All results were approved for release by the U.S. Census Bureau, authorization number CBDRB-FY22-CES018-010.}
\centering
\includegraphics[height=.8\textwidth,keepaspectratio]{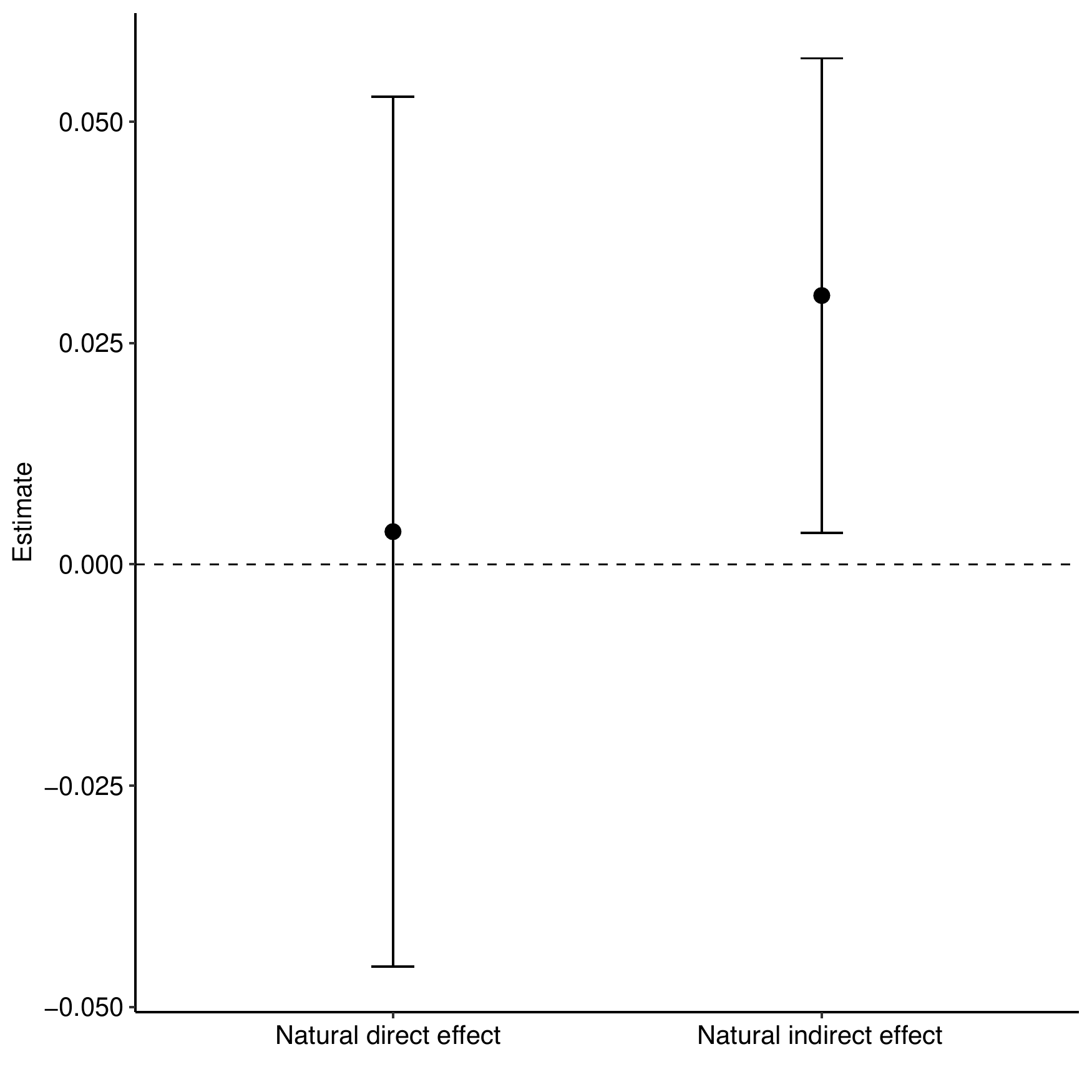}
\label{fig:examp}
\end{figure}

Figure \ref{fig:examp} shows the point estimates and 95\% confidence intervals for the natural direct effect and natural indirect effect. While the natural direct effect appears null, the natural indirect effect shows evidence of an unintended harmful path from being randomized to the housing voucher group on later risk of developing a psychiatric disorder operating through voucher-induced changes in the school and neighborhood environments and the instability of those environments, among boys. This indirect effect was estimated to contribute to a 3.04 percentage point increased risk of developing such a disorder (95\% CI: 0.36, 5.72 percentage points), which is in-line with previous estimates using population interventional indirect effects as the estimand \citep{rudolph2021helped}.

\section{Conclusion}
Although natural direct and indirect effects are not generally identifiable in the presence of treatment-induced confounders, assuming a monotonic relationship between the treatment and treatment-induced confounder achieves identifiability, as shown by \citet{tchetgen2014identification}. However, we are not aware of epidemiologists or other applied researchers using this identification strategy in practice. One reason could be that such scenarios are believed to be rare, niche circumstances. However, the monotonicity assumption is likely satisfied in any trial or intervention where the intervention assigns individuals to a treatment condition, but it is up to the individual whether or not to comply with the treatment they were assigned. Indeed, such study designs are common, and have been the setting for much mediation-related research \citep{vo2020conduct}. 
In this encouragement design setting, intervention take-up or adherence becomes the treatment-induced confounder, and one can then make the assumption that treatment assignment is monotonic with treatment take-up. If the remaining treatment-induced variables are treated as mediators, as we do in the applied example, then one can use our proposed one-step estimator. Our cross-fitted estimator is multiply robust and efficient, incorporates data-adaptive machine learning into model-fitting, and is available as an R package \url{link blinded for review}.

\bibliographystyle{plainnat}
\bibliography{refs}

\newpage
\begin{appendix}
\large{\textbf{Supplementary Materials for \titlepaper}}
\vspace{1cm}
\renewcommand{\thesection}{S\arabic{section}}
\renewcommand{\thesubsection}{S\arabic{subsection}}

\section{Proofs}
\begin{proof}
The proof proceeds as follows. Define $R_{z,z'}(\eta,\tilde\eta) = \theta_{z,z'}(\tilde\eta) - \theta_{z,z'}(\eta) +\E[D_{z,z'}(O;\tilde \eta)]$. This is equal to:
\begin{align}
R_{z,z'}(\eta,\tilde\eta)&=E\big\{[\tilde H_{Y,z,z'} - H_{Y,z,z'}][\E(Y\mid A,M,Z,W)-\tilde \E(Y\mid A,M,Z,W)]\big\}\notag\\ 
&+\E\left\{[\tilde H_{M,z,z'} - H_{M,z,z'}]\int \tilde \E(Y\mid a,M,z,W)[\dd \P(m\mid a', z', W) - \dd \tilde \P(m\mid a', z', W)]\right\}\notag\\ 
&+\E\big\{[\tilde H_{Z,z,z'} - H_{Z,z,z'}][\P(Z=1\mid A,W)-\tilde \P(Z=1\mid A,W)]\big\}\notag\\
&+\E\big\{H_{Y,z,z'}[\E(Y\mid A,M,Z,W)-\tilde \E(Y\mid A,M,Z,W)]\big\}\label{eq:term1}\\ 
&+\E\left\{H_{Z,z,z'}\int \tilde \E(Y\mid a,M,z,W)[\dd \P(m\mid a', z', W) - \dd \tilde \P(m\mid a', z', W)]\right\}\label{eq:term2}\\ 
&+\E\big\{H_{Z,z,z'}[\P(Z=1\mid A,W)-\tilde \P(Z=1\mid A,W)]\big\}\label{eq:term3}\\
&+\E\big\{\tilde H_{W,z,z'}-H_{W,z,z'}\big\}\label{eq:term4}.
\end{align}
It remains to show that the sum of terms (\ref{eq:term1})-(\ref{eq:term4}) is equal to $R^\star_{z,z'}(\eta,\tilde\eta)$. We show the proof for $(z,z')=(1,1)$, the proofs for the other terms follow similar arguments. Term (\ref{eq:term1}) is equal to 
\begin{align*}(\ref{eq:term1})&=\int H_Y(\tilde a,m,z,w)[\E(Y\mid a, m , 1, w)-\tilde\E(Y\mid a, m , 1, w)]\dd\P(\tilde a,m,z,w)\\
&=\int\frac{\one\{z=1,\tilde a=a\}}{\P(a'\mid w)}\frac{\P(a'\mid m, z, w)}{\P(\tilde a\mid m, z, w)}[\E(Y\mid a, m , 1, w)-\tilde\E(Y\mid a, m , 1, w)]\dd\P(\tilde a\mid m, z, w)\dd\P(m,z,w)\\
&=\int\frac{\one\{z=1\}}{\P(a'\mid w)}\P(a'\mid m, z, w)[\E(Y\mid a, m , 1, w)-\tilde\E(Y\mid a, m , 1, w)]\dd\P(m,z,w)\\
&=\int\frac{\one\{z=1\}}{\P(a'\mid w)}\frac{\P(a', m, z, w)}{\p(m,z,w)}[\E(Y\mid a, m , 1, w)-\tilde\E(Y\mid a, m , 1, w)]\dd\P(m,z,w)\\
&=\int\frac{\one\{z=1\}}{\P(a'\mid w)}\P(a', m, z, w)[\E(Y\mid a, m , 1, w)-\tilde\E(Y\mid a, m , 1, w)]\dd\nu(m,z,w)\\
&=\int\P(Z=1\mid a', w)[\E(Y\mid a, m , 1, w)-\tilde\E(Y\mid a, m , 1, w)]\dd\P(m\mid a', 1, w)\dd\P(w)\\\
\end{align*}
Term (\ref{eq:term2}) equals
\begin{align*}
    (\ref{eq:term2})&=\int H_M(\tilde a,z,w)\tilde\E(Y\mid a, m , 1, w)[\dd\P(m\mid a',1,w)-\dd\tilde\P(m\mid a', 1, w)]\dd\P(\tilde a,z,w)\\
    &=\int \frac{\one\{z=1,\tilde a = a'\}}{\P(a'\mid w)}\tilde\E(Y\mid a, m , 1, w)[\dd\P(m\mid a',1,w)-\dd\tilde\P(m\mid a', 1, w)]\dd\P(\tilde a,z,w)\\
    &=\int \P(Z=1\mid a',w)\tilde\E(Y\mid a, m , 1, w)[\dd\P(m\mid a',1,w)-\dd\tilde\P(m\mid a', 1, w)]\dd\P(w)\\
\end{align*}
Term (\ref{eq:term3}) equals
\begin{align*}
(\ref{eq:term3})&=\int H_Z(a,z,w)[\P(Z=1\mid a',w)-\tilde \P(Z=1\mid a', w)]\dd\P(a,w)\\    
&=\int \E(Y\mid a, m , 1, w)\dd\P(m\mid a',1,w)[\P(Z=1\mid a',w)-\tilde \P(Z=1\mid a', w)]\dd\P(w)\\    
\end{align*}
And term (\ref{eq:term4}) equals
\begin{multline*}
(\ref{eq:term4})=\int [\tilde\E(Y\mid a, m , 1, w)\dd\tilde\P(m\mid a',1,w)\tilde\P(Z=1\mid a',w)-\\\E(Y\mid a, m , 1, w)\dd\P(m\mid a',1,w)\P(Z=1\mid a',w)]\dd\P(w).
\end{multline*}
Factorizing these four terms gives the desired result.
\end{proof}

\section{Details of variables used in the empirical illustration}
Baseline covariates used:
 \begin{itemize}
     \item Adolescent characteristics: site (Boston, Chicago, LA, NYC), age, race/ethnicity (categorized as black, latino/Hispanic, white, other), number of family members (categorized as 2, 3, or 4+), someone from school asked to discuss problems the child had with schoolwork or behavior during the 2 years prior to baseline, child enrolled in special class for gifted and talented students.
     \item Adult household head characteristics included: high school graduate, marital status (never vs ever married), whether had been a teen parent, work status, receipt of AFDC/TANF, whether any family member has a disability.
     \item Neighborhood characteristics: felt neighborhood streets were unsafe at night; very dissatisfied with neighborhood; poverty level of neighborhood.
     \item Reported reasons for participating in MTO: to have access to better schools.
     \item Moving-related characteristics: moved more then 3 times during the 5 years prior to baseline, previous application for Section 8 voucher.
 \end{itemize}
 Mediator variables were duration-weighted (i.e., calculated as a weighted mean where weights were proportionate to the length of follow-up time (e.g., length of time that the youth attended each school, lived in each neighborhood, etc.)) in the 10-15 years between randomization and the final follow-up timepoint when the outcome was measured. These variables included:
 \begin{itemize}
     \item School rank
     \item School student-to-teacher ratio
     \item School's percent of students receiving free or reduced lunch
     \item Proportion of Title-I schools attended
     \item Number of moves
     \item Number of schools attended
     \item Number of school changes within the school year
     \item Whether or not the most recent school was in the same school district as baseline
     \item Neighborhood poverty
 \end{itemize}
 
\end{appendix}
\end{document}